\pdfoutput=1
\documentclass[journal]{IEEEtran}
\IEEEoverridecommandlockouts
\usepackage{lineno}
\usepackage{cite}
\usepackage{hyperref}
\usepackage{amsmath,amssymb,amsfonts}
\usepackage{amsmath}
\usepackage{amsthm}

\usepackage{graphicx}
\usepackage{textcomp}
\usepackage{xcolor}
\usepackage{graphicx}
\usepackage{float}
\usepackage{subfigure}
\usepackage{amsmath}
\usepackage{amsfonts,amssymb}
\usepackage{mathrsfs}
\usepackage{mathtools}
\usepackage{algorithm}
\usepackage{algorithmicx}
\usepackage{algpseudocode}
\usepackage{bm}
\usepackage{multirow}
\usepackage{array}
\usepackage{amssymb}
\usepackage{amsmath}
\usepackage{cite}
\usepackage{url}
\usepackage{xcolor}
\usepackage{cite,graphicx,amsmath,amssymb}
\usepackage{subfigure}
\usepackage{fancyhdr}
\usepackage{mdwmath}
\usepackage{mdwtab}
\usepackage{caption}
\usepackage{amsthm}
\usepackage{threeparttable}
\usepackage{setspace}
\usepackage{bm}
\usepackage{algorithm}
\usepackage{algpseudocode}
\usepackage{mathtools}
\usepackage{dsfont}
\usepackage{bbm}

\theoremstyle{definition}
\newtheorem{theorem}{Theorem}

\newtheorem{lemma}{Lemma}

\newtheorem{corollary}{Corollary}

\makeatletter
\newcommand{\biggg}{\bBigg@{3}}
\newcommand{\Biggg}{\bBigg@{3.5}}
\makeatother
\begin{document}
\title{Near-Field Communications: A Degree-of-Freedom Perspective}% New Looks at Interference in Communication Network (Or:
\author{Chongjun~Ouyang,~Yuanwei~Liu,~Xingqi~Zhang,~and~Lajos~Hanzo
\thanks{C. Ouyang is with the School of Information and Communication Engineering, Beijing University of Posts and Telecommunications, Beijing, 100876, China (e-mail: DragonAim@bupt.edu.cn).}
\thanks{Y. Liu is with the School of Electronic Engineering and Computer Science, Queen Mary University of London, London, E1 4NS, U.K. (e-mail: yuanwei.liu@qmul.ac.uk).}
\thanks{Q. Zhang is with the Department of Electrical and Computer Engineering, University of Alberta, Edmonton, T6G 2H5, Canada (e-mail: xingqi.zhang@ualberta.ca).}
\thanks{L. Hanzo is with the School of Electronics and Computer Science, University of Southampton, Southampton, SO17 1BJ, U.K. (e-mail: lh@ecs.soton.ac.uk).}}

\maketitle

\begin{abstract}
Multiple-antenna technologies are advancing towards large-scale aperture sizes and extremely high frequencies, leading to the emergence of near-field communications (NFC) in future wireless systems. To this context, we investigate the degree of freedom (DoF) in near-field multiple-input multiple-output (MIMO) systems. We consider both spatially discrete (SPD) antennas and continuous aperture (CAP) antennas. Additionally, we explore three important DoF-related performance metrics and examine their relationships with the classic DoF. Numerical results demonstrate the benefits of NFC over far-field communications (FFC) in terms of providing increased spatial DoFs. We also identify promising research directions for NFC from a DoF perspective.
\end{abstract}

\section{Introduction}

%The electromagnetic (EM) radiation field emitted by antennas can be categorized into two main regions: the far-field and the radiation near-field. The boundary that separates these regions is known as the Rayleigh distance, which is determined by the product of the square of the array aperture and the carrier frequency \cite{Liu2023}. In the far-field region, which extends beyond the Rayleigh distance, EM waves exhibit distinct propagation characteristics compared to the near-field region within the Rayleigh distance. Specifically, the far-field EM field can be effectively approximated by planar waves, while the near-field EM field requires precise modeling using spherical waves \cite{Liu2023}.

The electromagnetic (EM) radiation field emitted by antennas is divided into two regions: the far-field and the radiation near-field. The Rayleigh distance, determined by the product of the array aperture's square and the carrier frequency, serves as the boundary between these regions \cite{Liu2023}. In the far-field region, beyond the Rayleigh distance, EM waves exhibit different propagation characteristics compared to the near-field region within it. Planar waves effectively approximate the far-field EM field, while the near-field EM field requires precise modeling using spherical waves \cite{Liu2023}.

Limited by the size of antenna arrays and the operating frequency bands, the Rayleigh distance in current cellular systems typically spans only a few meters, making the near-field effects negligible. Thus, existing cellular communications predominantly rely on theories and techniques from far-field communications (FFC). However, with the rapid advances of wireless technology, next-generation wireless communications rely on extremely large-scale antenna arrays and higher frequencies to cater for the ever-increasing thirst for communication services \cite{Cui2023}. In these advanced scenarios, near-field communications (NFC) can extend over longer distances, surpassing the conventional proximity range. The deployment of massive antenna arrays and the utilization of high-frequency bands allow NFC to be effective at distances of hundreds of meters, thereby opening up novel opportunities for the development of NFC theories and techniques \cite{Liu2023,Cui2023}.

In the realm of wireless communications, the degree of freedom (DoF) concept has emerged as a crucial framework for understanding the capabilities and potential of different communication systems \cite{Tse2005}. Briefly, the DoF provides insights into the number of independent signal dimensions that can be exploited for conveying information in a wireless channel. While traditional FFC have been extensively studied within this context, the unique physical properties of NFC exhibit distinct characteristics that necessitate a fresh exploration of DoF.

The adoption of a DoF perspective in NFC is motivated by several factors. Firstly, NFC offers increased DoFs, which represents a significant advantage over FFC. By understanding the DoF characteristics of NFC systems, we can unveil the superior data capacity and transmission capabilities of NFC compared to FFC. Secondly, characterizing the DoF in NFC assists in optimizing the system parameters, such as the antenna configurations and transmission strategies, leading to improved overall performance. Thirdly, adopting a DoF perspective facilitates the development of communication protocols and algorithms specifically tailored for NFC environments, resulting in enhanced reliability, coverage, and throughput. Although there are some studies analyzing NFC's DoF \cite{Liu2023}, this field is still in its infancy.

\begin{figure}[!t]
\centering
\setlength{\abovecaptionskip}{0pt}
\includegraphics[height=0.35\textwidth]{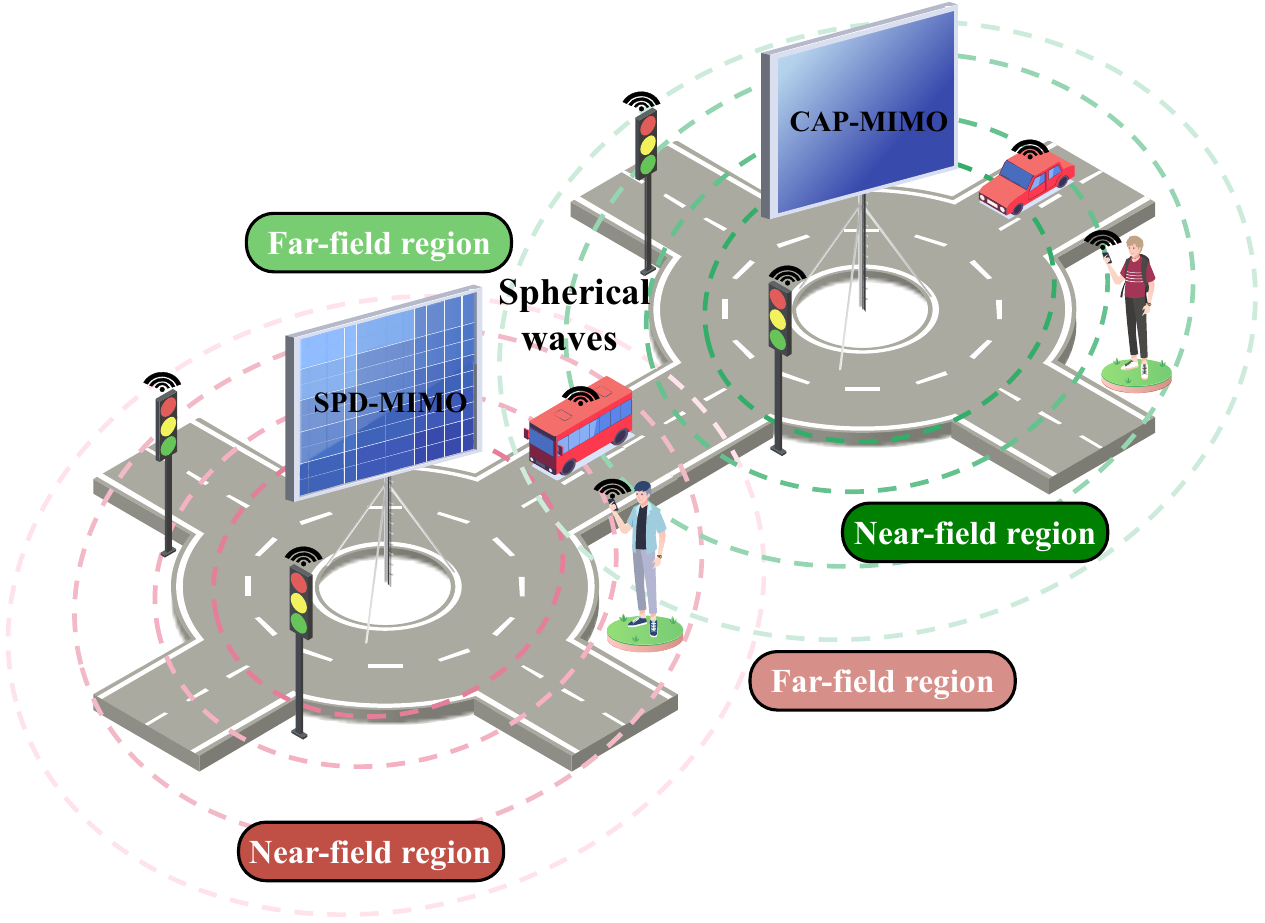}
\caption{Illustration of near-field communications.}
\label{System_Model}
\vspace{-10pt}
\end{figure}

Hence, we aim for the critical appraisal of NFC and its DoF. Our focus is on point-to-point multiple-input multiple-output (MIMO) channels under line-of-sight (LoS) propagation, as illustrated in {\figurename} {\ref{System_Model}}. This emphasis arises from the anticipation that future NFC will operate at high frequencies, leading to a prevalence of LoS communication associated with limited multi-path effects. We commence by exploring the DoFs achieved in near-field MIMO by spatially discrete antennas (SPD-MIMO). Subsequently, we extend our analysis to the near-field MIMO supported by continuous aperture antennas (CAP-MIMO). Utilizing numerical simulations, we demonstrate the superiority of NFC over FFC concerning its DoF and establish connections between the DoF and effective DoF (EDoF). Finally, future research ideas are discussed.

\section{DoFs Achieved in SPD-MIMO}
In practical implementations of NFC, a viable approach is to equip the transceiver with an extensive antenna array comprising a large number of SPD patch antennas. %These antennas are closely spaced, usually with a distance less than half the wavelength.
In this section, we will delve into a comprehensive analysis of the achievable DoFs in near-field SPD-MIMO. %Moreover, we will conduct a comparative study of the achievable DoF in both the near-field and far-field scenarios.

\subsection{Calculation of the DoF}\label{Section_Calculation_SPD}
\subsubsection{${\mathsf{DoF}}$}
In the context of SPD-MIMO, the overall channel response can be represented as a matrix $\mathbf{H}$ having dimensions of $N_{\rm{r}}\times N_{\rm{t}}$, where $N_{\rm{r}}$ denotes the number of receive antennas and $N_{\rm{t}}$ represents the number of transmit antennas. By applying the singular value decomposition (SVD) to this channel matrix, the SPD-MIMO channel can be effectively decomposed into multiple independent single-input single-output (SISO) sub-channels that operate in parallel without mutual interference. Mathematically, the number of positive singular values or the rank of the correlation matrix $\mathbf{H}{\mathbf{H}}^{\mathsf{H}}$ corresponds to the number of sub-channels having a non-zero signal-to-noise ratio (SNR). Each of these sub-channels accommodates an independent communication mode within the MIMO channel. The total number of communication modes is referred to as the \emph{spatial DoF} of the channel, denoted as $\mathsf{DoF}$. On the other hand, for a MIMO Gaussian channel, the capacity growth rate can be shown to be ${\mathsf{DoF}}\cdot\log_2({\mathsf{SNR}})$ at high SNR. Therefore, the DoF is also termed as the \emph{high-SNR slope} or maximum \emph{multiplexing gain} (relative to a SISO channel) \cite{Liu2023}.

Given a channel matrix $\mathbf{H}$, the spatial DoFs are inherently limited and cannot exceed the minimum value between $N_{\rm{r}}$ and $N_{\rm{t}}$. In a far-field MIMO LoS channel, only a single incident angle is available due to the almost parallel planar-wave propagation. Consequently, the channel matrix is of rank-$1$, resulting in a very limited DoF, namely $1$. By contrast, within the near-field region, the spherical waves exhibit different phase-shifts and power levels for each link. This diversity leads to a higher rank for the MIMO channel matrix and subsequently a higher DoF compared to the far-field scenario. Notably, if the SPD antennas are well separated, the achievable DoFs for the near-field MIMO LoS channel can approach the minimum value between $N_{\rm{r}}$ and $N_{\rm{t}}$. \emph{This signifies that spatial multiplexing can be supported even in the absence of a rich scattering environment, which is a significant advantage of NFC.}

\begin{figure*}[!t]
    \centering
    \subfigbottomskip=0pt
%	\subfigcapskip=-5pt
    \setlength{\abovecaptionskip}{0pt}
    \subfigure[SPD-MIMO. $\{{\bm\phi}_n\}_{n=1}^{{\mathsf{EDoF}}_1}$ and $\{{\bm\psi}_n\}_{n=1}^{{\mathsf{EDoF}}_1}$ are the right and left singular vectors of ${\mathbf{H}}$ corresponding to the dominant ${{\mathsf{EDoF}}_1}$ singular values $\{{\sigma}_n\}_{n=1}^{{\mathsf{EDoF}}_1}$, $\mathbf{x}$ is the transmitted signal vector, and $\mathbf{y}$ is the received signal vector.]
    {
        \includegraphics[width=0.7\textwidth]{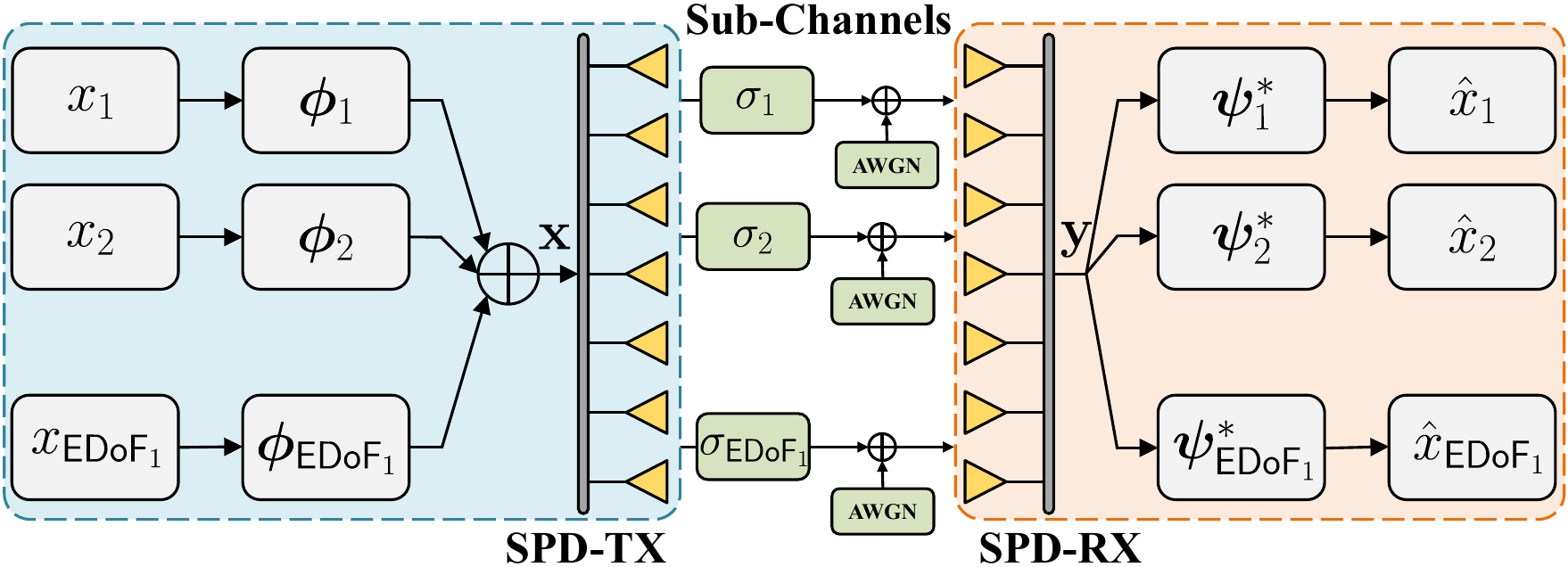}
	   \label{SVD_SPD}	
    }
   \subfigure[CAP-MIMO. $\{{\bm\phi}_n(\cdot)\}_{n=1}^{{\mathsf{EDoF}}_1}$ and $\{{\bm\psi}_n(\cdot)\}_{n=1}^{{\mathsf{EDoF}}_1}$ are the right and left singular functions of Green's function corresponding to the dominant ${{\mathsf{EDoF}}_1}$ singular values $\{{\sigma}_n\}_{n=1}^{{\mathsf{EDoF}}_1}$, $\mathbf{J}(\cdot)$ is the continuous distribution of source currents, $\mathbf{E}(\cdot)$ is the electric radiation field generated at the receiver, $\mathbf{s}$ is the source point within the CAP surface of the transmitter, and $\mathbf{r}$ is the field point within the CAP surface of the receiver.]
    {
        \includegraphics[width=0.7\textwidth]{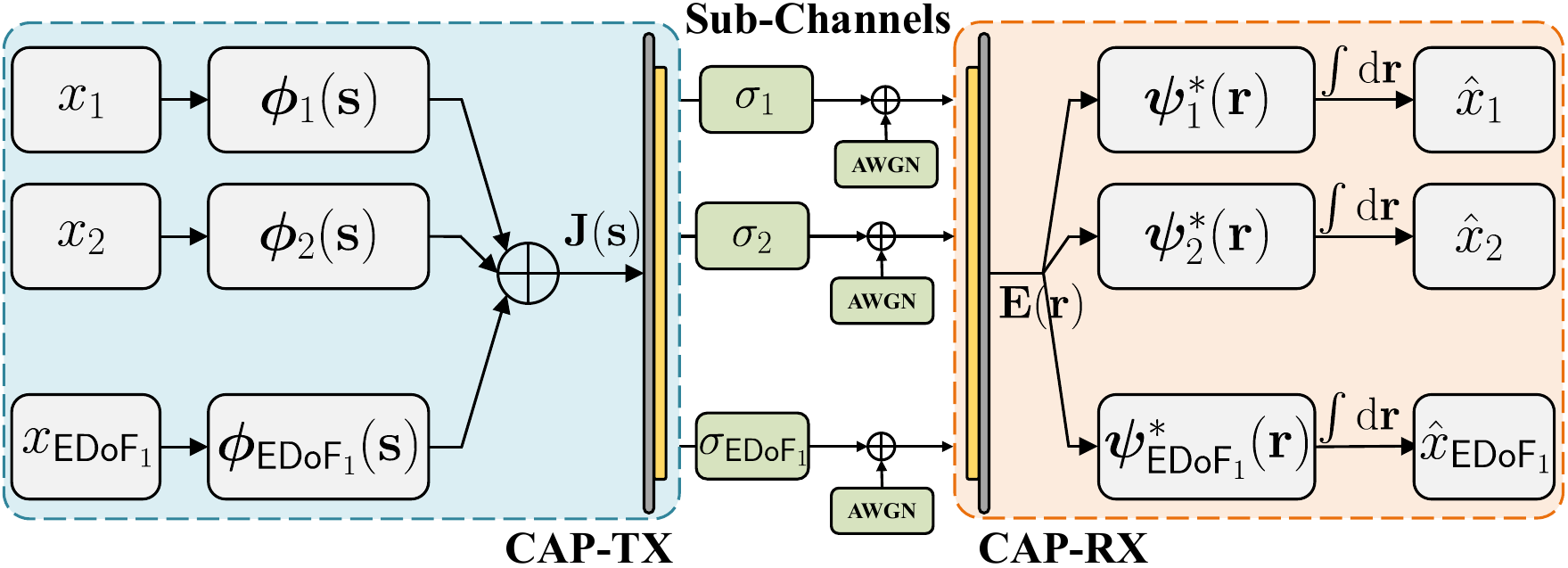}
	   \label{SVD_CAP}	
    }
%\centering
%\setlength{\abovecaptionskip}{0pt}
%\includegraphics[height=0.3\textwidth]{SVD_SPD.eps}
\caption{Communication architecture based on orthogonal parallel sub-channels for MIMO NFC. In this figure, $\{x_n\}_{n=1}^{{\mathsf{EDoF}}_1}$ are the transmitted symbols, $\{\hat{x}_n\}_{n=1}^{{\mathsf{EDoF}}_1}$ are the received symbols, $\{{\sigma}_n\}_{n=1}^{{\mathsf{EDoF}}_1}$ are the dominant singular values of the MIMO channel (channel matrix or Green's function), and AWGN stands for additive white Gaussian noise.}
\label{figure_architecture}
\vspace{-10pt}
\end{figure*}

\subsubsection{${\mathsf{EDoF}}_1$}
The aforementioned arguments suggest that employing a high number of antennas constitutes an effective technique of increasing the DoFs in NFC. By reducing the antenna spacing within a fixed aperture size, the number of spatial DoFs can be expanded. It is worth noting that when two antennas are in each other's close proximity, the waves they generate at the receiver antenna array become nearly identical. Consequently, these two antennas become indistinguishable at the receiver. This limitation should be considered as it could restrict the potential increase in channel capacity, when a large number of transceiving antennas are incorporated into a fixed aperture. This limitation has been theoretically demonstrated in \cite{Miller2000,Dardari2020}.

To augment our exposition, we represent the ordered positive singular values of matrix $\mathbf{H}$ as $\sigma_1\geq\ldots\geq\sigma_{\mathsf{DoF}}$. Miller \cite{Miller2000} demonstrated by employing prolate spheroidal wave functions that for small values of $n$, the $\sigma_n$ values fall off slowly until they reach a critical threshold, beyond which they decay rapidly. This critical threshold is termed as the ``\emph{effective degree of freedom (EDoF)}'', denoted as ${\mathsf{EDoF}}_1$. Moreover, this phenomenon becomes more prominent as the number of transceiving antennas increases. These findings indicate that although harnessing more antennas can lead to an increased number of independent sub-channels, only the dominant ${\mathsf{EDoF}}_1$ ones can be effectively utilized for supporting reliable communications.

%Furthermore, for a sufficiently large number of antennas, the work presented in \cite{Miller2000} concludes that for NFC systems facilitated by a linear antenna array, the upper limit of ${\mathsf{EDoF}}_1$ is proportional to the product of the transmitter and receiver lengths and inversely proportional to the distance between the transmitter and receiver. Similarly, for NFC systems facilitated by a planar array, the upper limit of ${\mathsf{EDoF}}_1$ is proportional to the product of the transmitter and receiver areas and inversely proportional to the square of the link distance.

%Furthermore, for a large number of antennas, \cite{Miller2000} concludes that in NFC systems with linear antenna arrays, the upper limit of ${\mathsf{EDoF}}_1$ is proportional to the product of transmitter and receiver lengths and inversely proportional to the distance between them. Similarly, for NFC systems with planar arrays, the upper limit of ${\mathsf{EDoF}}_1$ is proportional to the product of transmitter and receiver areas and inversely proportional to the square of the link distance.

Furthermore, for a large number of antennas, Miller \cite{Miller2000} concludes that the upper limit of ${\mathsf{EDoF}}_1$ is proportional to the product of transmitter and receiver areas and it is inversely proportional to the link distance. These findings are derived using the uniform spherical wave (USW) model described in \cite[Eqn. (35)]{Liu2023}. The USW model is applicable in the near-field region, where the communication distance exceeds the uniform-power distance, exhibiting uniform channel gains and non-linear phase-shifts. However, it is important to note that as the link distance becomes comparable to the transceiver sizes (i.e., NFC within the uniform-power distance), the accuracy of the USW model and the EDoF derived in \cite{Miller2000} diminishes. To address this, Dardari introduced a more general formula for ${\mathsf{EDoF}}_1$ based on 2D sampling theory arguments for the non-uniform spherical wave (NUSW) model of \cite{Dardari2020}. Although this formula may present tractability challenges, again, it reveals that the upper limit of ${\mathsf{EDoF}}_1$ is proportional to the product of the transmitter and receiver areas, while it is inversely proportional to the link distance. These improvements enhance our understanding of EDoF in NFC systems.

%The findings presented in \cite{Miller2000} are derived from the application of the uniform spherical wave (USW) model, as described in \cite[Eqn. (35)]{Liu2023}. The USW model is applicable to the near-field region where the communication distance exceeds the uniform-power distance. This region exhibits uniform channel gains and non-linear phase-shifts. For a more comprehensive understanding, we refer readers to our tutorial paper \cite{Liu2023} for further details on the USW model. However, it is important to note that when the link distance becomes comparable to the physical sizes of the transceivers, i.e., NFC within the uniform-power distance, the accuracy of the USW model and the derived EDoF in \cite{Miller2000} diminishes. To address this limitation, Dardari introduced a more general formula for ${\mathsf{EDoF}}_1$ based on the 2D sampling theory arguments for the non-uniform spherical wave (NUSW) model \cite[Eqn. (31)]{Dardari2020}. Although this formula may present challenges in terms of tractability, it reveals that the upper limit of ${\mathsf{EDoF}}_1$ is proportionate to the product of the transmitter and receiver areas while inversely proportional to the link distance. These improvements enhance our understanding of the EDoF in NFC systems.

In summary, the conclusions drawn from \cite{Miller2000} and \cite{Dardari2020} suggest that the number of dominant communication modes and channel capacity
can be enhanced in two primary means: increasing the aperture size and reducing the communication distance. Remarkably, these strategies align with the commonly employed techniques for supporting NFC, emphasizing the superior spatial EDoF capabilities of NFC systems.

\subsection{Exploitation of the DoF}
To fully utilize the increased DoFs or EDoFs (i.e., ${\mathsf{EDoF}}_1$) offered by near-field SPD-MIMO, it is crucial to apply SVD to the channel matrix $\mathbf{H}$. This allows for the identification of the right and left singular vectors corresponding to the dominant ${\mathsf{EDoF}}_1$ singular values. To further optimize the achievable channel capacity, the water-filling algorithm can be utilized for judiciously sharing the power among the ${\mathsf{EDoF}}_1$ parallel sub-channels. {\figurename} {\ref{SVD_SPD}} illustrates the detailed architecture that outlines the exploitation of DoF in NFC relying on SPD antennas.

%It is important to acknowledge that the computational complexity associated with SVD increases cubically with the number of transceiving antennas. As the number of antennas deployed becomes greater, it becomes crucial to identify low-complexity SVD solutions or approximations. This ensures the practical feasibility of implementing the proposed techniques.

\begin{table*}[!t]
\caption{Summary of DoF-related metrics for MIMO NFC supported by SPD antennas.}
\label{table_spd}
\small
\centering
\begin{threeparttable}
\resizebox{0.95\textwidth}{!}{
\begin{tabular}{|ll|r|rrr|}
\hline
\multicolumn{2}{|l|}{\multirow{2}{*}{Metric}}                                                                              & Degree of Freedom: DoF                                                                                         & \multicolumn{3}{r|}{Effective Degree of Freedom: EDoF}                                                                                                                                                                                                                                                                                        \\ \cline{3-6}
\multicolumn{2}{|r|}{}                                                                                                     & ${\mathsf{DoF}}$                                                                                         & \multicolumn{1}{r|}{${\mathsf{EDoF}}_1$}                                                                                    & \multicolumn{1}{r|}{${\mathsf{EDoF}}_2$}                                                                                          & ${\mathsf{EDoF}}_3$                                                                         \\ \hline
\multicolumn{2}{|l|}{Definition}                                                                                           & \cite{Tse2005}                                                                             & \multicolumn{1}{r|}{\cite{Miller2000}}                                                                         & \multicolumn{1}{r|}{\cite{Muharemovic2008}}                                                                                & \cite{Shiu2000}                                                               \\ \hline
\multicolumn{2}{|l|}{Values Range}                                                                                           & $\in{\mathbb{Z}}^{+}$, $[1,N_{\min}]$                                                                             & \multicolumn{1}{r|}{$\in{\mathbb{Z}}^{+}$, $[1,N_{\min}]$}                                                                         & \multicolumn{1}{r|}{$\in{\mathbb{R}}^{+}$, $[1,N_{\min}]$}                                                                                & $\in{\mathbb{R}}^{+}$, $(0,N_{\min}]$                                                               \\ \hline
\multicolumn{2}{|l|}{SNR Ranges}                                                                                           & High-SNR region                                                                             & \multicolumn{1}{r|}{Low\&Medium-SNR region}                                                                         & \multicolumn{1}{r|}{Low-SNR region}                                                                                & All SNR ranges                                                               \\ \hline
\multicolumn{2}{|l|}{\begin{tabular}[l]{@{}l@{}}Relation with \\ Sub-Channels\end{tabular}}                                & \begin{tabular}[r]{@{}r@{}}Number of sun-channels \\ with a non-zero SNR\end{tabular} & \multicolumn{1}{r|}{\begin{tabular}[r]{@{}r@{}}Number of\\ dominant sub-channels\end{tabular}} & \multicolumn{1}{r|}{\begin{tabular}[r]{@{}r@{}}No direct relation with \\ the number of sub-channels\end{tabular}} & \begin{tabular}[r]{@{}r@{}}Number of \\ equivalent sub-channels\end{tabular} \\ \hline
\multicolumn{1}{|l|}{\multirow{3}{*}{\begin{tabular}[l]{@{}l@{}}SPD-MIMO\\Fae-Field\\ (Calculation)\end{tabular}}} & LoS                   & $1$                                                                                           & \multicolumn{1}{r|}{$1$}                                                                                       & \multicolumn{1}{r|}{$1$}                                                                                             & $\leq1$                                                                            \\ \cline{2-6}
\multicolumn{1}{|l|}{}                                                                             & \multirow{2}{*}{NLoS} & Rank of $\mathbf{H}{\mathbf{H}}^{\mathsf{H}}$, $\geq1$                                                                                    & \multicolumn{1}{r|}{Obtained from SVD of $\mathbf{H}$, $\geq1$}                                                                                & \multicolumn{1}{r|}{$({\mathsf{tr}}({\mathbf{HH}}^{\mathsf{H}})/\lVert{\mathbf{HH}}^{\mathsf{H}}\rVert_{\rm{F}})^2$ \cite{Muharemovic2008}, $\geq1$}                                                                                        & $\left.\frac{{\rm{d}}}{{\rm{d}}\delta}C({\mathsf{SNR}}\cdot2^{\delta})\right|_{\delta=0}$ \cite{Shiu2000}, $\leq N_{\min}$                                                                         \\ \cline{3-6}
\multicolumn{1}{|l|}{}                                                                             &                       & Upper bound: $N_{\min}$                                                                           & \multicolumn{1}{r|}{Upper bound: $N_{\min}$}                                                                       & \multicolumn{1}{r|}{Upper bound: $N_{\min}$}                                                                             & Upper bound: $N_{\min}$                                                            \\ \hline
\multicolumn{1}{|l|}{\multirow{4}{*}{\begin{tabular}[l]{@{}l@{}}SPD-MIMO\\ Near-Field\\ (Calculation)\end{tabular}}} & \multirow{2}{*}{LoS}  & Rank of $\mathbf{H}{\mathbf{H}}^{\mathsf{H}}$, $\geq1$                                                                   & \multicolumn{1}{r|}{Obtained from SVD of $\mathbf{H}$, $\geq1$}                                                               & \multicolumn{1}{r|}{$({\mathsf{tr}}({\mathbf{HH}}^{\mathsf{H}})/\lVert{\mathbf{HH}}^{\mathsf{H}}\rVert_{\rm{F}})^2$, \cite[Eqn. (8)]{Xie2023}, $\geq1$}                                                                                        & $\left.\frac{{\rm{d}}}{{\rm{d}}\delta}C({\mathsf{SNR}}\cdot2^{\delta})\right|_{\delta=0}$, $\leq N_{\min}$                                                                         \\ \cline{3-6}
\multicolumn{1}{|l|}{}                                                                             &                       & Upper bound: $N_{\min}$                                                                           & \multicolumn{1}{r|}{Upper limit: $\propto A_{\rm{t}}A_{\rm{r}}$, $\propto{d^{-2}}$, \cite{Miller2000,Dardari2020}}                                                                    & \multicolumn{1}{r|}{Upper limit: \cite[Eqn. (18)]{Xie2023}, $\propto{d^{-1}}$}                                                                                 & Upper bound: $N_{\min}$                                                            \\ \cline{2-6}
\multicolumn{1}{|l|}{}                                                                             & \multirow{2}{*}{NLoS} & Rank of $\mathbf{H}{\mathbf{H}}^{\mathsf{H}}$, $\geq1$                                                                                    & \multicolumn{1}{r|}{Obtained from SVD of $\mathbf{H}$, $\geq1$}                                                                                & \multicolumn{1}{r|}{$({\mathsf{tr}}({\mathbf{HH}}^{\mathsf{H}})/\lVert{\mathbf{HH}}^{\mathsf{H}}\rVert_{\rm{F}})^2$, $\geq1$}                                                                                        & $\left.\frac{{\rm{d}}}{{\rm{d}}\delta}C({\mathsf{SNR}}\cdot2^{\delta})\right|_{\delta=0}$, $\leq N_{\min}$                                                                         \\ \cline{3-6}
\multicolumn{1}{|l|}{}                                                                             &                       & Upper bound: $N_{\min}$                                                                           & \multicolumn{1}{r|}{Upper limit: $\propto A_{\rm{t}}A_{\rm{r}}$, \cite{Poon2005,Pizzo2023}}                                                                   & \multicolumn{1}{r|}{Upper bound: $N_{\min}$}                                                                             & Upper bound: $N_{\min}$                                                            \\ \hline
\end{tabular}}
\begin{tablenotes}
\footnotesize
\item[*] $A_{{\rm{t}}/{\rm{r}}}$ is the effective aperture size of the transmitter/receiver
\item[**] $N_{\min}$ is the minimum value between $N_{\rm{r}}$ and $N_{\rm{t}}$
\item[***] $d$ is the link distance between the transmitter and receiver
\end{tablenotes}
\end{threeparttable}
\end{table*}

\subsection{Discussion and Outlook}
\subsubsection{MIMO NLoS Channel}
%The preceding discussions in Section \ref{Section_Calculation_SPD} have primarily focused on MIMO LoS channels.
%However, it is worth noting that examining the number of DoFs achieved in near-field NLoS MIMO channels still holds theoretical significance.
The DoF of MIMO NLoS channels is influenced by the geometrical distribution of scatterers. In a rich scattering environment, the MIMO channel can achieve full rank for both the near-field and far-field regions due to the random phase shifts introduced by scatterers. As a result, the achievable DoFs in MIMO LoS channels may approach the minimum value between the numbers of receive and transmit antennas. When a large number of transceiving antennas are employed, the authors of \cite{Poon2005} and \cite{Pizzo2023} have demonstrated by leveraging sampling theory that the upper limit of ${\mathsf{EDoF}}_1$ is directly proportional to the effective aperture of the transceivers.
%In summary, while the focus has been on MIMO LoS channels, studying the DoFs achieved in Near-Field NLoS MIMO channels is still of great theoretical importance, especially considering the influence of scatterer geometry. Additionally, in environments with a significant number of transceiving antennas, the upper limit of EDoF1 can be linked to the effective aperture of the transceivers, as shown in [4] and [5].

For SPD-MIMO, the exact values of $\mathsf{DoF}$ and ${\mathsf{EDoF}}_1$ can be obtained from the SVD of the channel matrix $\mathbf{H}$ for both LoS and NLoS channels. However, obtaining tractable closed-form expressions for these two performance metrics remains challenging. To address this, previous studies have investigated the upper limit of ${\mathsf{EDoF}}_1$ under various channel conditions by considering the asymptotic scenario of a large number of transceiving antennas \cite{Dardari2020,Pizzo2023,Poon2005,Miller2000}. These elegant expressions are derived using Green's function model, which may appear impervious to newcomers, who are experts in other fields. This leads to an important question: \emph{Can there be DoF-related metrics that evaluate NFC performance in a non-asymptotic manner in closed form?} The answer is affirmative, and the following parts provide the details of these metrics.
\subsubsection{${\mathsf{EDoF}}_2$}
Recently, some researchers have introduced an alternative metric to assess NFC performance, also termed as the ``\emph{effective degree of freedom (EDoF)}'', which is given by $({\mathsf{tr}}({\mathbf{HH}}^{\mathsf{H}})/\lVert{\mathbf{HH}}^{\mathsf{H}}\rVert_{\rm{F}})^2$ and denoted as ${\mathsf{EDoF}}_2$. ${\mathsf{EDoF}}_2$ can be readily calculated for any arbitrary channel matrix, regardless of whether the system operates in near- or far-field regions, and under LoS or NLoS propagations. As an example, let us consider the LoS channel. In far-field LoS MIMO, the channel matrix has a rank of $1$, and hence ${\mathsf{EDoF}}_2$ becomes $1$. Conversely, for near-field LoS MIMO, ${\mathsf{EDoF}}_2$ falls between $1$ and ${\mathsf{DoF}}$, and it is also proportional to the number of transceiving antennas \cite{Xie2023}. The upper limit of ${\mathsf{EDoF}}_2$ is obtained for near-field LoS MIMO by letting the number of antennas approach infinity, demonstrating its inverse proportionality to the link distance. The results in \cite{Xie2023} indicate that the near-field effect can enhance ${\mathsf{EDoF}}_2$. Several studies have claimed, without any justifications, that ${\mathsf{EDoF}}_2$ represents the equivalent number of sub-channels, as depicted in {\figurename} {\ref{SVD_SPD}}, and can be employed for evaluating the NFC performance \cite{Xie2023}. However, it is crucial to note that these statements lack mathematical rigor and may lead to misinterpretations of the actual meaning and implications of ${\mathsf{EDoF}}_2$.

The concept of ${\mathsf{EDoF}}_2$ was originally introduced by Muharemovic \emph{et al.} \cite{Muharemovic2008}, who built upon Verd\'{u}'s previous work \cite{Verdu2002} to approximate the MIMO channel capacity as ${\mathsf{EDoF}}_2\cdot[\log_2(\frac{E_b}{N_0})-\log_2({\frac{E_b}{N_0}}_{\min})]$ in the low-SNR regime. Here, $\frac{E_b}{N_0}$ represents the bit energy over noise power spectral density, and ${\frac{E_b}{N_0}}_{\min}$ is the minimum value required for reliable communications. Additionally, $\frac{E_b}{N_0}$ is determined by the product of the channel capacity and the SNR \cite[Eqn. (14)]{Verdu2002}. By considering the insights gleaned from \cite{Muharemovic2008} and \cite{Verdu2002}, it becomes evident that ${\mathsf{EDoF}}_2$ possesses a distinct physical interpretation when compared to ${\mathsf{EDoF}}_1$ and ${\mathsf{DoF}}$. Generally, the value of ${\mathsf{EDoF}}_2$ is not directly associated with the number of dominant sub-channels depicted in {\figurename} {\ref{SVD_SPD}}. However, an exception occurs when the dominant sub-channels have nearly identical channel gains, i.e., $\sigma_1\approx\ldots\approx\sigma_{\mathsf{EDoF}_1}\gg \sigma_{\mathsf{EDoF}_1+1}>\ldots>\sigma_{\mathsf{DoF}}$. In such cases, ${\mathsf{EDoF}}_1$ can be approximately represented by the value of ${\mathsf{EDoF}}_2$. Our numerical results in Section \ref{Section:Numerical} suggest that this scenario can happen in certain LoS channels. Nonetheless, this approximation remains heuristic, and its generality lacks mathematical rigor.

To summarize, ${\mathsf{EDoF}}_2$ serves as a significant performance metric for NFC in the low-SNR region, yet it cannot be simply interpreted as the equivalent number of sub-channels. Its significance and interpretation are different from those of ${\mathsf{EDoF}}_1$ and ${\mathsf{DoF}}$. Hence, it is important to discern its distinct role in NFC.

\subsubsection{${\mathsf{EDoF}}_3$}
To fully harness the spatial DoFs offered by NFC MIMO, it is desirable to operate the system in the high-SNR region. In such scenarios, the channel capacity should exhibit roughly linear growth vs. the ${\mathsf{DoF}}$ or ${\mathsf{EDoF}}_1$, given a fixed transmit power. However, achieving this high SNR condition may not always be feasible in practical settings. In recognition of this fact, Shiu \emph{et al.} \cite{Shiu2000} introduced an alternative metric, also termed as the ``\emph{effective degree of freedom (EDoF)}'', which represents the number of equivalent sub-channels actively participating in conveying information under specific operating conditions. For clarity, we refer to this metric as ${\mathsf{EDoF}}_3$.

In a SISO channel, a $G$-fold increase in transmit power leads to a capacity increase of $\log_2{G}$ bps/Hz at high SNRs. If a system is equivalent to ${\mathsf{EDoF}}_3$ SISO channels in parallel, the overall system capacity should increase by ${\mathsf{EDoF}}_3\cdot\log_2{G}$ bps/Hz when the transmit power is multiplied by a factor of $G$. To formally define ${\mathsf{EDoF}}_3\cdot\log_2{G}$, Shiu \emph{et al.} \cite{Shiu2000} express it as $\left.\frac{{\rm{d}}}{{\rm{d}}\delta}C({\mathsf{SNR}}\cdot2^{\delta})\right|_{\delta=0}$, where $C({\mathsf{SNR}})$ represents the MIMO channel capacity at a given SNR. It is important to note that $C(\cdot)$ can refer to the instantaneous capacity, outage capacity, or ergodic capacity, making the expression of ${\mathsf{EDoF}}_3$ applicable to arbitrary channel matrices, regardless of whether the system operates in the near- or far-field regions, and under LoS or NLoS propagations. Let us consider the LoS channel as an example. In far-field LoS MIMO, the channel matrix has a rank of $1$, leading to ${\mathsf{EDoF}}_3$ being no larger than $1$. Conversely, for near-field LoS MIMO, ${\mathsf{EDoF}}_3$ could exceed $1$ \cite{Shiu2000}. Observe from this comparison that the near-field effect can improve ${\mathsf{EDoF}}_3$.

Essentially, ${\mathsf{EDoF}}_3$ describes the number of equivalent SISO sub-channels at a given SNR, making it a valuable performance indicator for NFC in different SNR scenarios.

%\subsubsection{Comparison Among ${\mathsf{DoF}}$, ${\mathsf{EDoF}}_1$, ${\mathsf{EDoF}}_2$, and ${\mathsf{EDoF}}_3$}
%By comparing ${\mathsf{DoF}}$, ${\mathsf{EDoF}}_1$, ${\mathsf{EDoF}}_2$, and ${\mathsf{EDoF}}_3$, we observe the following results.
%\begin{itemize}
%\item ${\mathsf{DoF}}$ characterizes the number of sun-channels with a non-zero SNR, ${\mathsf{EDoF}}_1$ characterizes the number of dominant sub-channels, ${\mathsf{EDoF}}_3$ characterizes the number of equivalent sub-channels, while ${\mathsf{EDoF}}_2$ is not directly associated with the number of sub-channels.
%\item ${\mathsf{DoF}}$ is the pre-log factors of the channel capacity in the high-SNR regime (in terms of ${\mathsf{SNR}}$), ${\mathsf{EDoF}}_2$ is the pre-log factor in the low-SNR regime (in terms of $\frac{E_b}{N_0}$),  ${\mathsf{EDoF}}_3$ is the pre-log factor in all SNR ranges (in terms of ${\mathsf{SNR}}$).
%\item ${\mathsf{EDoF}}_3$ is a function of the SNR, while ${\mathsf{DoF}}$, ${\mathsf{EDoF}}_1$, and ${\mathsf{EDoF}}_2$ are not.
%\end{itemize}

%Taken together, the four DoF-related metrics, i.e., ${\mathsf{DoF}}$, ${\mathsf{EDoF}}_1$, ${\mathsf{EDoF}}_2$, and ${\mathsf{EDoF}}_3$, possess different physical meanings and scopes of application. As such, they should be appropriately utilized based on the practical demands of NFC.

\subsubsection{Summary and Outlook}
A detailed comparison among ${\mathsf{DoF}}$, ${\mathsf{EDoF}}_1$, ${\mathsf{EDoF}}_2$, and ${\mathsf{EDoF}}_3$ is summarized in Table \ref{table_spd}. Taken together, these four DoF-related metrics possess different physical meanings and scopes of application. As such, they should be appropriately utilized based on the practical demands of NFC.
%The findings presented in Table \ref{table_spd} demonstrate that the minimum value between $N_{\rm{r}}$ and $N_{\rm{t}}$ serves as a general upper bound for the DoF and EDoFs. Additionally, several upper limits of the EDoFs are obtained for SPD-MIMO with a large number of transceiving antennas, and these limits are influenced by the aperture size or link distance.
While the existing results have been primarily focused on ${\mathsf{EDoF}}_1$, it is crucial to develop a comprehensive mathematical framework for calculating the upper limits of ${\mathsf{EDoF}}_2$ and ${\mathsf{EDoF}}_3$ under both LoS and NLoS scenarios. This avenue represents a potential direction for future research. %The increased DoFs resulting from the near-field effect offer opportunities for exploiting enhanced spatial multiplexing gains and improving the overall system capacity. However, the discussions in this section are primarily focused on the point-to-point MIMO channel. Further research is warranted to delve into the fundamental analysis and applications of spatial DoFs in multiuser MIMO NFC, as this area holds promising prospects for in-depth exploration and investigation.

%Near-field communications offer a unique advantage for discrete ELAA MIMO systems by increasing the available DoF. We explore the mechanisms through which NFC enables the exploitation of additional spatial dimensions for information transfer. By leveraging the close proximity of the ELAA and the receiving devices, we can achieve enhanced spatial multiplexing gains and improve the overall system capacity.
%
%We present theoretical analysis and simulation results to demonstrate the increased DoF achievable with NFC for discrete ELAA MIMO systems, highlighting the potential for high-dimensional signal transmission.
%
%The increased DoF in near-field communications for discrete ELAA presents opportunities for improving network throughput. We discuss strategies for effectively utilizing the additional spatial dimensions to enhance data rates and system performance. Techniques such as advanced precoding, beamforming, and interference management can be employed to exploit the increased DoF and mitigate the effects of near-field channel characteristics.

%We showcase examples and case studies that illustrate the impact of leveraging DoF to achieve higher network throughput in these advanced communication systems.

\section{DoFs Achieved in CAP-MIMO}
Utilizing CAP antennas presents a promising technique of improving the performance of MIMO systems having limited apertures.
%, often referred to as holographic MIMO.
In contrast to SPD-MIMOs, which involve a large number of discrete antennas having specific spacing, CAP-MIMO adopts an infinite number of antennas with infinitesimal spacing. %Consequently, the CAP surface can be seen as a limit of the SPD antenna array. As a result, the achievable EDoF in near-field CAP-MIMO essentially represent the upper limit of EDoF in SPD-MIMO. In this section, we present a detailed investigation into the spatial DoFs in near-field CAP-MIMO.
This section investigates the spatial DoFs in near-field CAP-MIMO.
\subsection{Calculation of the DoF}
We consider a scenario where both the transmitter and receiver are equipped with CAP antennas, which is analogous to the MIMO setup for SPD antennas. However, in contrast to the SPD antenna array that delivers finite-dimensional signal vectors, the CAP surface supports a continuous distribution of source currents within the transmitting aperture, giving rise to the generation of an electric radiation field at the receiver aperture. The spatial channel impulse response between any two points on the transceiving surfaces is described by Green's function, which connects the transmitter's current distribution and the receiver's electric field via a spatial integral. Green's function accurately models the EM characteristics in free space and effectively represents the channel response between the transceivers, akin to the channel matrix for SPD-MIMOs.
\subsubsection{${\mathsf{DoF}}$}
Based on the above considerations, the spatial CAP-MIMO channel can be decomposed into a series of parallel SISO sub-channels by finding the equivalent ``SVD'' of Green's function \cite[Eqn. (27)]{Miller2000}. The resultant equivalent ``left singular vectors'' and ``right singular vectors'' form two complete sets of orthogonal basis functions, one for the transmitter's aperture and the other for the receiver's aperture. The resultant equivalent ``singular values'' correspond to the channel gains of the decomposed sub-channels. Alternatively, these ``singular values'' can be obtained through the eigenvalue decomposition of the Hermitian kernel of Green's function (analogous to the correlation matrix ${\mathbf{H}}{\mathbf{H}}^{\mathsf{H}}$ for SPD antennas); see \cite[Eqn. (42)]{Miller2000} and \cite[Section \uppercase\expandafter{\romannumeral2-C}]{Liu2023} for more details. The number of non-zero ``singular values'' of Green's function, or equivalently, the non-zero eigenvalues of its kernel, is defined as the DoF, denoted as ${\mathsf{DoF}}$. The DoF also signifies the number of SISO sub-channels at a non-zero SNR, each of which supports an independent communication mode within the entire system.

As noted in \cite{Miller2000}, the far-field LoS CAP-MIMO can support a maximum of one communication mode. Consequently, the DoF of far-field LoS CAP-MIMO is limited to $1$. However, in the case of near-field LoS MIMO, the DoF has the potential to approach infinity due to the associated spherical wave propagation \cite{Miller2000}. Therefore, we may conclude that the near-field effect significantly enhances the spatial DoFs for CAP-MIMO.
\subsubsection{${\mathsf{EDoF}}_1$}
The near-field CAP-MIMO system has the remarkable ability to support infinitely many communication modes. However, it is crucial to recognize that only those modes having significant channel gains can be effectively utilized to convey information. The total number of these effective communication modes is known as the EDoF, i.e., ${\mathsf{EDoF}}_1$. Several methods have been proposed to determine or approximate the value of ${\mathsf{EDoF}}_1$, such as analyzing the eigenvalues of the kernel of Green's function \cite{Miller2000}, employing sampling theory \cite{Poon2005,Dardari2020,Pizzo2023}, utilizing diffraction theory \cite{Xu2023}, or leveraging Landau's theorem \cite{Pizzo2022}.

Prior research has demonstrated that for near-field LoS CAP-MIMO, the value of ${\mathsf{EDoF}}_1$ is directly proportional to the product of the transmitter and receiver areas while being inversely proportional to the link distance \cite{Poon2005,Dardari2020,Pizzo2023}. On the other hand, for far-field LoS CAP-MIMO, the value of ${\mathsf{EDoF}}_1$ is limited to $1$. These findings highlight the superiority of NFC in terms of enhancing the spatial DoFs.
%, particularly in near-field scenarios where the potential for multiple effective communication modes significantly increases.
\begin{table*}[!h]
\caption{Summary of DoF-related metrics for MIMO NFC supported by CAP antennas.}
\label{table_cap}
\small
\centering
\begin{threeparttable}
\resizebox{0.8\textwidth}{!}{
\begin{tabular}{|ll|r|rrr|}
\hline
\multicolumn{2}{|l|}{\multirow{2}{*}{Metric}}                                                             & Degree of Freedom: DoF                                                                                             & \multicolumn{3}{r|}{Effective Degree of Freedom: EDoF}                                                                                                                                                                                                                                                                                        \\ \cline{3-6}
\multicolumn{2}{|l|}{}                                                                                    & $\mathsf{DoF}$                                                                                             & \multicolumn{1}{r|}{$\mathsf{EDoF}_1$}                                                                                    & \multicolumn{1}{r|}{$\mathsf{EDoF}_2$}                                                                                          & $\mathsf{EDoF}_3$                                                                         \\ \hline
\multicolumn{2}{|l|}{Definition}                                                                                           & \cite{Tse2005}                                                                             & \multicolumn{1}{r|}{\cite{Miller2000}}                                                                         & \multicolumn{1}{r|}{\cite{Muharemovic2008}}                                                                                & \cite{Shiu2000}                                                               \\ \hline
\multicolumn{2}{|l|}{Values Range}                                                                                           & $\in{\mathbb{Z}}^{+}$, $[1,\infty)$                                                                             & \multicolumn{1}{r|}{$\in{\mathbb{Z}}^{+}$, $[1,\infty)$}                                                                         & \multicolumn{1}{r|}{$\in{\mathbb{R}}^{+}$, $[1,\infty)$}                                                                                & $\in{\mathbb{R}}^{+}$, $(0,\infty]$
\\ \hline
\multicolumn{2}{|l|}{SNR Ranges}                                                                          & High-SNR region                                                                                 & \multicolumn{1}{r|}{Low\&Medium-SNR region}                                                                         & \multicolumn{1}{r|}{Unknown}                                                                                & All SNR ranges                                                               \\ \hline
\multicolumn{2}{|l|}{\begin{tabular}[l]{@{}l@{}}Relation with \\ Sub-Channels\end{tabular}}               & \begin{tabular}[l]{@{}r@{}}Number of sub-channels \\ with a non-zero SNR\end{tabular}     & \multicolumn{1}{r|}{\begin{tabular}[r]{@{}r@{}}Number of\\ dominant sub-channels\end{tabular}} & \multicolumn{1}{r|}{\begin{tabular}[r]{@{}r@{}}No direct relation with \\ the number of sub-channels\end{tabular}} & \begin{tabular}[r]{@{}r@{}}Number of \\equivalent sub-channels\end{tabular} \\ \hline
\multicolumn{1}{|l|}{\multirow{2}{*}{\begin{tabular}[l]{@{}l@{}}CAP-MIMO \\Far-Field \\(Calculation)\end{tabular}}} & LoS  & $1$                                                                                               & \multicolumn{1}{r|}{$1$}                                                                                       & \multicolumn{1}{r|}{$1$}                                                                                             & $\leq1$                                                                            \\ \cline{2-6}
\multicolumn{1}{|l|}{}                                                                             & NLoS & \begin{tabular}[r]{@{}r@{}}Obtained from solving the \\ eigenvalue problem, $\geq1$\end{tabular} & \multicolumn{1}{r|}{\begin{tabular}[r]{@{}r@{}}Obtained from solving the \\ eigenvalue problem, $\geq1$\end{tabular}}                                                                                & \multicolumn{1}{r|}{\begin{tabular}[r]{@{}r@{}}The exact expression\\ is unknown, $\geq1$\end{tabular}}                                                                                        & $\left.\frac{{\rm{d}}}{{\rm{d}}\delta}C({\mathsf{SNR}}\cdot2^{\delta})\right|_{\delta=0}$                                                                         \\ \hline
\multicolumn{1}{|l|}{\multirow{2}{*}{\begin{tabular}[l]{@{}l@{}}CAP-MIMO \\Near-Field\\ (Calculation)\end{tabular}}} & LoS  & \begin{tabular}[r]{@{}r@{}}Obtained from solving the \\ eigenvalue problem, $\geq1$\end{tabular} & \multicolumn{1}{r|}{$\propto A_{\rm{t}}A_{\rm{r}}$, $\propto{d^{-2}}$, $\geq1$, \cite{Miller2000,Dardari2020}}                                                                            & \multicolumn{1}{r|}{$\propto d^{-1}$, $\geq1$, \cite{Jiang2023,Xie2023}}                                                                                  & $\left.\frac{{\rm{d}}}{{\rm{d}}\delta}C({\mathsf{SNR}}\cdot2^{\delta})\right|_{\delta=0}$                                                                         \\ \cline{2-6}
\multicolumn{1}{|l|}{}                                                                             & NLoS & \begin{tabular}[r]{@{}r@{}}Obtained from solving the \\ eigenvalue problem, $\geq1$\end{tabular} & \multicolumn{1}{r|}{$\propto A_{\rm{t}}A_{\rm{r}}$, \cite{Poon2005,Pizzo2023}}                                                                            & \multicolumn{1}{r|}{\begin{tabular}[r]{@{}r@{}}The exact expression\\ is unknown, $\geq1$\end{tabular}}                                                                                        & $\left.\frac{{\rm{d}}}{{\rm{d}}\delta}C({\mathsf{SNR}}\cdot2^{\delta})\right|_{\delta=0}$                                                                         \\ \hline
\end{tabular}}
\begin{tablenotes}
\footnotesize
\item[**] $A_{{\rm{t}}/{\rm{r}}}$ is the effective aperture size of the transmitter/receiver, $d$ is the link distance between the transmitter and receiver
\end{tablenotes}
\end{threeparttable}
\end{table*}
\subsection{Exploitation of the DoF}
To fully exploit the increased EDoFs offered by near-field CAP-MIMO, it becomes essential to determine the left and right singular functions of Green's function and their associated singular values. This task involves solving the eigenvalue problem for the Hermitian kernel \cite{Liu2023}. A potential architecture for the CAP-MIMO is illustrated in {\figurename} {\ref{SVD_CAP}}, which closely resembles that of SPD-MIMO. However, it is important to acknowledge that the computational complexity associated with solving the eigenvalue problem for CAP-MIMO is significantly higher than that for SPD-MIMO. Additionally, the architecture depicted in {\figurename} {\ref{SVD_CAP}} requires the use of infinitely many radio-frequency chains.

\subsection{Discussion and Outlook}
\subsubsection{MIMO NLoS Channel}
The DoFs of near-field CAP-MIMO have also been investigated in the context of NLoS propagation. In \cite{Poon2005} and \cite{Pizzo2023}, the authors explored various scattering environments and utilized sampling theory to analyze the EDoF. Their findings revealed that $\mathsf{EDoF}_1$ of NLoS CAP-MIMO is higher than $1$ in both the near-field and far-field regions. Moreover, they demonstrated that increasing the effective aperture of the transceivers can lead to further improvements of $\mathsf{EDoF}_1$.
\subsubsection{${\mathsf{EDoF}}_2$}
The concept of $\mathsf{EDoF}_2$ has been extended to CAP-MIMO channels upon replacing the channel matrix by Green's function \cite[Eqn. (8)]{Jiang2023}. Closed-form formulas of $\mathsf{EDoF}_2$ have been derived for near-field CAP-MIMO \cite{Xie2023,Jiang2023}, specifically for the LoS channel. The analysis reveals that while $\mathsf{EDoF}_2$ of FFC is limited to $1$, $\mathsf{EDoF}_2$ of NFC is inversely proportional to the link distance. These findings underscore the advantage of NFC in terms of $\mathsf{EDoF}_2$. However, it is essential to acknowledge that there are currently no studies proving that the channel capacity of CAP-MIMO satisfies ${\mathsf{EDoF}}_2\cdot[\log_2(\frac{E_b}{N_0})-\log_2({\frac{E_b}{N_0}}_{\min})]$ in the low-SNR regime. As a result, ${\mathsf{EDoF}}_2$ remains a heuristic concept for CAP-MIMO, lacking precise physical interpretations. Further research is needed to establish a more rigorous and practical understanding of ${\mathsf{EDoF}}_2$ in the context of CAP-MIMO.
\subsubsection{${\mathsf{EDoF}}_3$}
The concept of ${\mathsf{EDoF}}_3$ is also applicable to CAP-MIMO. It is evident that ${\mathsf{EDoF}}_3$ of a far-field LoS channel cannot exceed $1$, while for near-field LoS CAP-MIMO, ${\mathsf{EDoF}}_3$ can be higher than $1$. However, it is important to note that due to the lack of closed-form expressions for the channel capacity of CAP-MIMO, calculating the exact value of ${\mathsf{EDoF}}_3$ for near-field CAP-MIMO becomes intractable \cite{Wan2023}. Therefore, further investigations are required to address this aspect and gain a deeper understanding of ${\mathsf{EDoF}}_3$ in the context of near-field CAP-MIMO.

\begin{figure*}[!t]
    \centering
    \subfigbottomskip=0pt
%	\subfigcapskip=-5pt
    \setlength{\abovecaptionskip}{0pt}
    \subfigure[$\mathsf{EDoF}_1$.]
    {
        \includegraphics[height=0.24\textwidth]{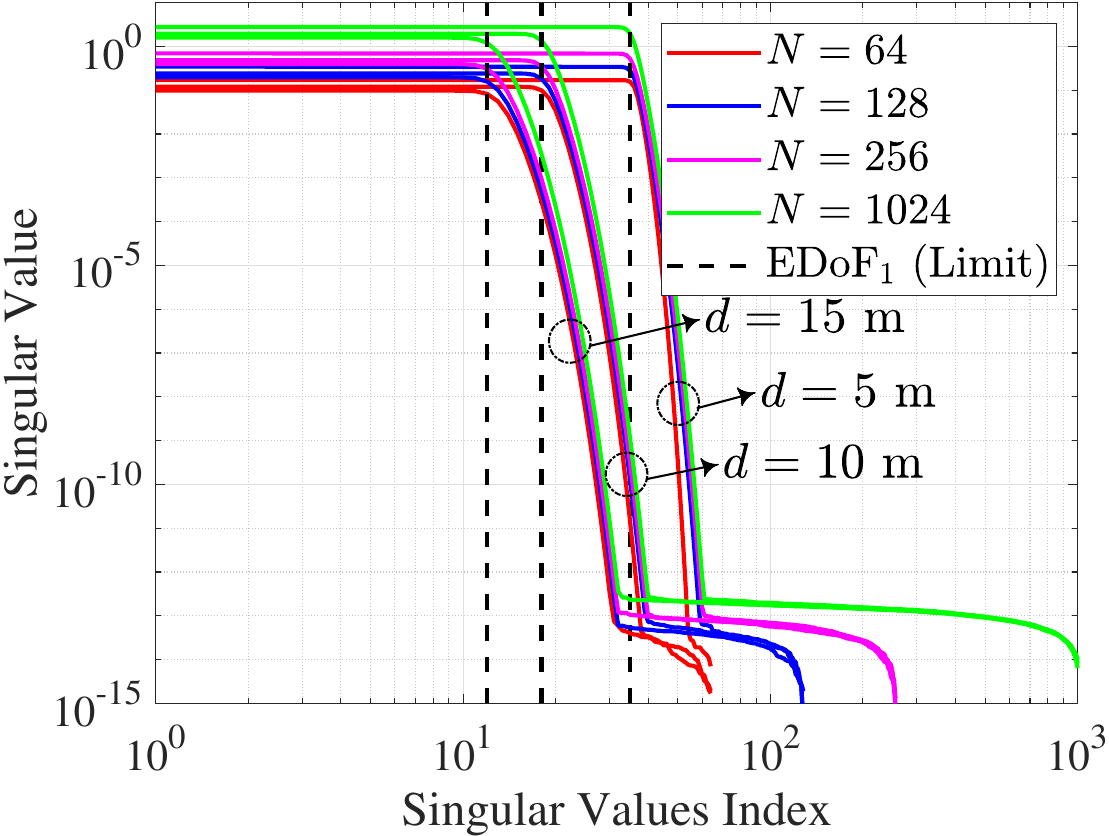}
	   \label{figure:EDoF1_SPD}	
    }
   \subfigure[$\mathsf{EDoF}_2$.]
    {
        \includegraphics[height=0.24\textwidth]{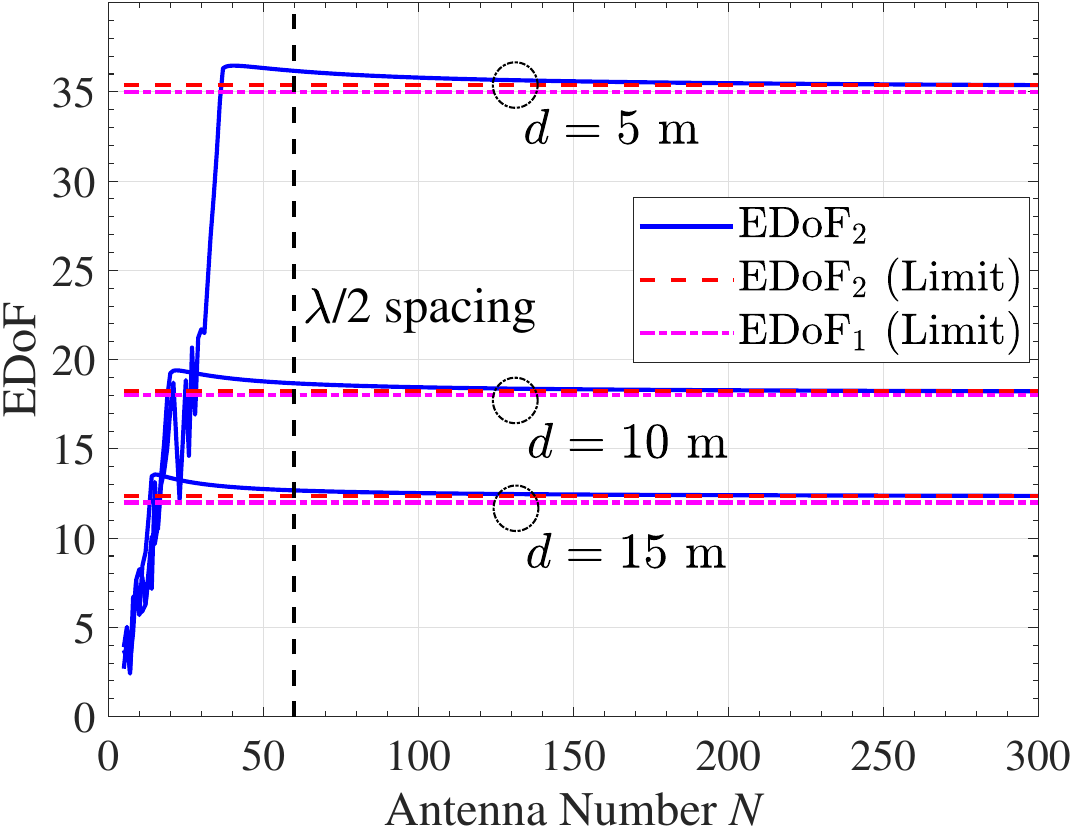}
	   \label{figure:EDoF2_SPD}	
    }
    \subfigure[$\mathsf{EDoF}_3$.]
    {
        \includegraphics[height=0.24\textwidth]{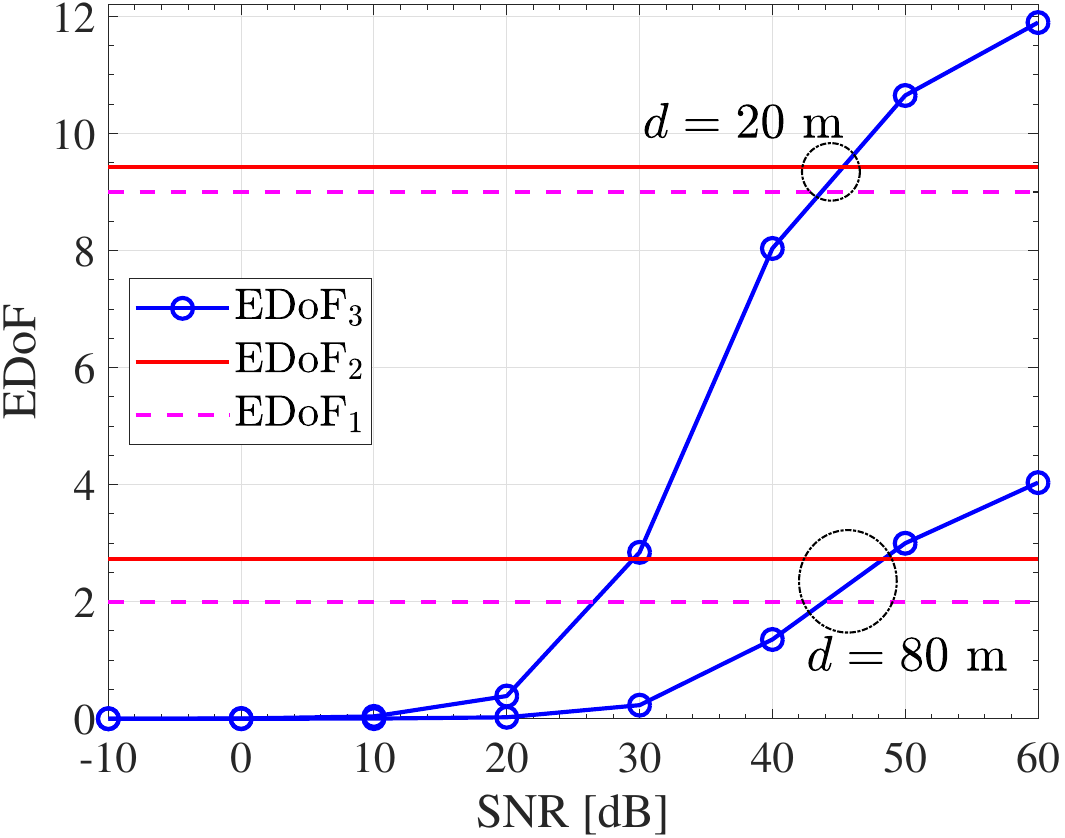}
	   \label{figure:EDoF3_SPD}	
    }
\caption{Illustration of EDoFs in SPD-MIMO LoS channels, where both transmitter and receiver are equipped with uniform linear arrays (ULAs), each containing $N$ antennas, and the system operates at a frequency of $28$ GHz (with a corresponding wavelength of $\lambda=1$ cm). The ULAs have an aperture size of $1.37$ m. The center of the transmitter is located at the origin of a three-dimensional plane, while the center of the receiver is at $(0,d,0)$ with $d$ denoting the link distance. The ULAs face each other and are parallel to the $z$-axis.}
\label{figure_SPD_EDoF}
\vspace{-10pt}
\end{figure*}
\subsubsection{Summary and Outlook}

A detailed comparison of ${\mathsf{DoF}}$, ${\mathsf{EDoF}}_1$, ${\mathsf{EDoF}}_2$, and ${\mathsf{EDoF}}_3$ is summarized in Table \ref{table_cap}. The results presented in Table \ref{table_cap} primarily pertain to point-to-point CAP-MIMO channels. However, investigating the spatial DoFs introduced by the near-field effect in a multiuser CAP-MIMO setup holds both theoretical and practical significance. As previously mentioned, the practical implementation of near-field CAP-MIMO is computationally intractable. Therefore, it is imperative to explore practical and scalable techniques of CAP-MIMO implementations.
%, enabling us to fully leverage the advantages of increased EDoFs in near-field scenarios.

\section{Numerical Results}\label{Section:Numerical}
In this section, we explore the enhanced DoFs and EDoFs offered by MIMO NFC through computer simulations in LoS channel scenarios.

\subsection{SPD-MIMO}
{\figurename} {\ref{figure_SPD_EDoF}} illustrates the DoFs and EDoFs in SPD-MIMO, showcasing the increased DoFs provided by the near-field effect. Specifically, in {\figurename} {\ref{figure:EDoF1_SPD}}, we present the singular values of the MIMO channel matrix for different link distances and numbers of antennas. Notably, the DoF of NFC is significantly higher than the value achieved by FFC, surpassing the single DoF threshold. As shown, the singular values exhibit a slow decline until they reach a critical threshold, after which they decrease rapidly. The number of dominant singular values defines the ${\mathsf{EDoF}}_1$. From {\figurename} {\ref{figure:EDoF1_SPD}}, we can infer that as the number of antennas increases, the singular values, and thus the channel gains of the decomposed sub-channels, experience slight improvements, with ${\mathsf{EDoF}}_1$ converging rapidly to its upper limit (as calculated in \cite{Miller2000}). Additionally, it is noteworthy that a shorter link distance results in a higher ${\mathsf{EDoF}}_1$ \cite{Miller2000}, showcasing the superiority of NFC in terms of DoF enhancement.

{\figurename} {\ref{figure:EDoF2_SPD}} presents the plot of ${\mathsf{EDoF}}_2$ for SPD-MIMO in the near-field. We observe that as the number of antennas increases, ${\mathsf{EDoF}}_2$ of SPD-MIMO converges to its limit, which is equivalent to ${\mathsf{EDoF}}_2$ of CAP-MIMO \cite{Xie2023}. This convergence occurs more rapidly for higher link distances. Remarkably, as depicted in the graph, SPD-MIMO having half-wavelength antenna spacing can achieve nearly the same ${\mathsf{EDoF}}_2$ as CAP-MIMO. The results in {\figurename} {\ref{figure:EDoF1_SPD}} indicate that the singular values of our system satisfy $\sigma_1\approx\ldots\approx\sigma_{\mathsf{EDoF}_1}\gg \sigma_{\mathsf{EDoF}_1+1}>\ldots>\sigma_{\mathsf{DoF}}$, where ${\mathsf{EDoF}}_1$ can be approximated by the value of ${\mathsf{EDoF}}_2$. This observation aligns with the findings from {\figurename} {\ref{figure:EDoF2_SPD}}.

In {\figurename} {\ref{figure:EDoF3_SPD}}, we plot ${\mathsf{EDoF}}_3$ as a function of the SNR. Observe that reducing the link distance enhances ${\mathsf{EDoF}}_3$, further validating the use of the near-field effect to improve channel capacity. Additionally, we note that in the high-SNR regime, ${\mathsf{EDoF}}_3$ can exceed ${\mathsf{EDoF}}_1$ and ${\mathsf{EDoF}}_2$. This phenomenon arises because the non-dominant sub-channels can also support reliable communications, when sufficient transmit power resources are available.

%Overall, the results presented in {\figurename} {\ref{figure_SPD_EDoF}} provide valuable insights into the relationships between DoFs and EDoFs in SPD-MIMO and demonstrate the significant potential of the near-field effect for enhancing MIMO NFC performance.
\begin{figure*}[!t]
    \centering
    \subfigbottomskip=0pt
%	\subfigcapskip=-5pt
    \setlength{\abovecaptionskip}{0pt}
    \subfigure[$\mathsf{EDoF}_1$.]
    {
        \includegraphics[width=0.3\textwidth]{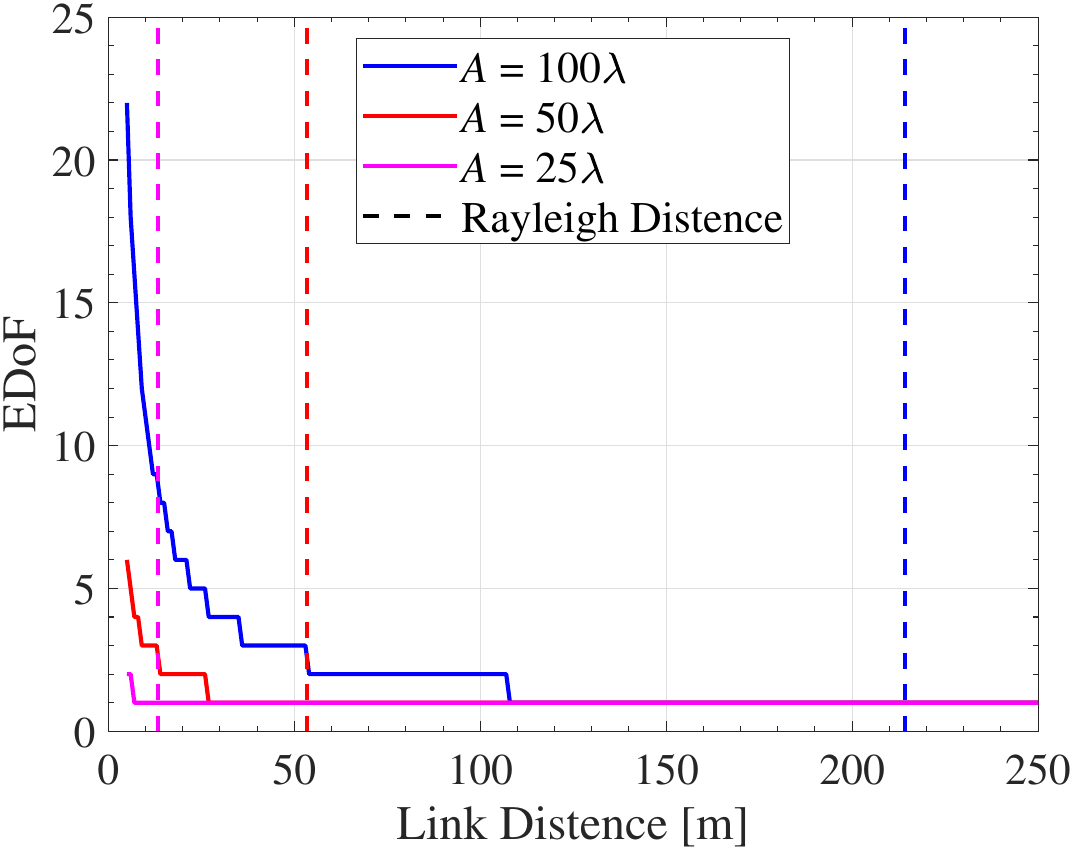}
	   \label{figure:EDoF1_CAP}	
    }
   \subfigure[$\mathsf{EDoF}_2$.]
    {
        \includegraphics[width=0.3\textwidth]{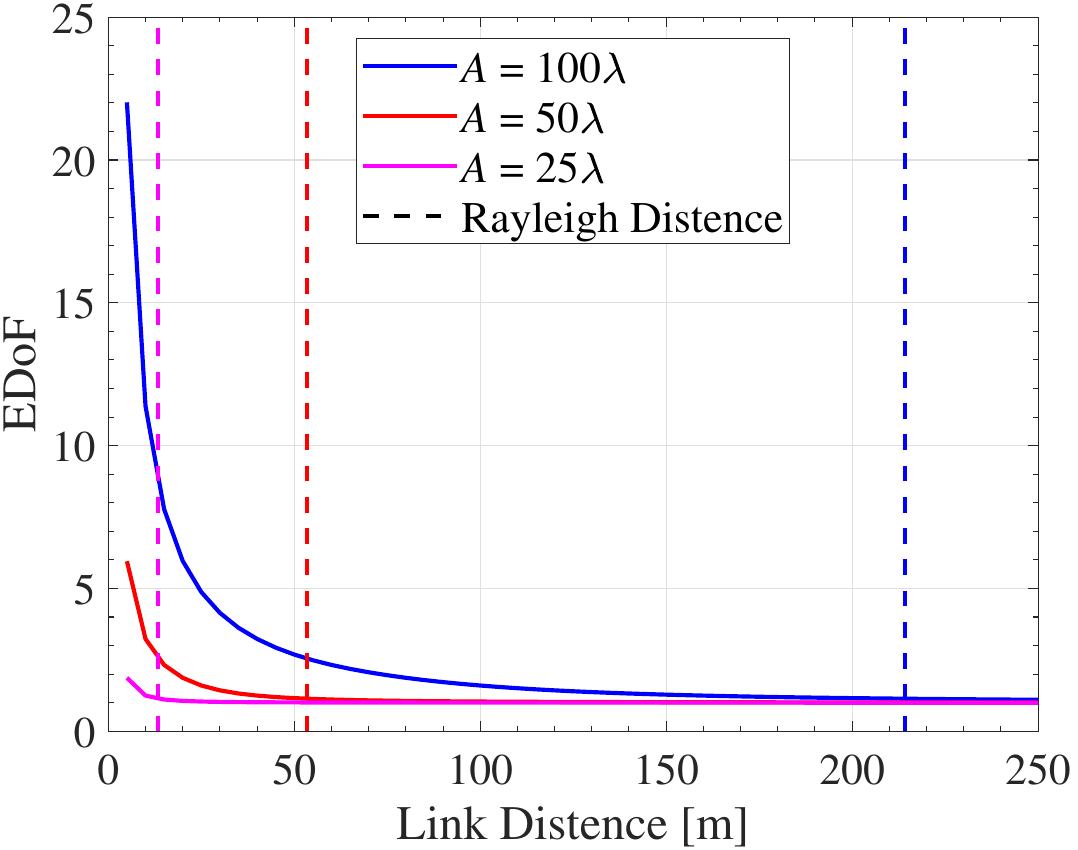}
	   \label{figure:EDoF2_CAP}	
    }
\caption{Illustration of EDoFs in CAP-MIMO LoS channels, where the transmitter and receiver are equipped with continuous linear arrays of the same aperture size $A$, and the system operates at a frequency of $28$ GHz (with a corresponding wavelength of $\lambda=1$ cm). The center of the transmitter is located at the origin of a three-dimensional plane, while the center of the receiver is at $(0,d,0)$ with $d$ denoting the link distance. The linear arrays face each other and are parallel to the $z$-axis.}
\label{figure_CAP_EDoF}
\vspace{-10pt}
\end{figure*}

\subsection{CAP-MIMO}
{\figurename} {\ref{figure_CAP_EDoF}} presents an analysis of the DoFs in CAP-MIMO systems. Due to the computational complexity associated with calculating the channel capacity of CAP-MIMO \cite{Wan2023}, we focus on illustrating ${\mathsf{EDoF}}_1$ and ${\mathsf{EDoF}}_2$ in this figure, and the numerical results for ${\mathsf{EDoF}}_3$ are omitted.

In {\figurename} {\ref{figure:EDoF1_CAP}} and {\figurename} {\ref{figure:EDoF2_CAP}}, we showcase ${\mathsf{EDoF}}_1$ and ${\mathsf{EDoF}}_2$ as functions of the link distance, respectively. To differentiate between the near-field and far-field regions, we mark the Rayleigh distance in both graphs. The figures demonstrate that both ${\mathsf{EDoF}}_1$ and ${\mathsf{EDoF}}_2$ can be enhanced by either increasing the aperture sizes of the transceivers or reducing the link distance. These strategies align with commonly employed techniques for supporting NFC. A notable observation from the comparison of {\figurename} {\ref{figure:EDoF1_CAP}} and {\figurename} {\ref{figure:EDoF2_CAP}} is that the curves for ${\mathsf{EDoF}}_1$ follow similar trends to those of ${\mathsf{EDoF}}_2$, corroborating the findings from {\figurename} {\ref{figure:EDoF2_SPD}}.

The numerical results presented in {\figurename} {\ref{figure_SPD_EDoF}} and {\figurename} {\ref{figure_CAP_EDoF}} collectively underscore the substantial impact of near-field effects on augmenting the DoFs in MIMO systems. These findings contribute valuable insights to the understanding and design of NFC technologies.
\section{Conclusion and Promising Research Directions}
In this article, we have conducted an in-depth investigation into the performance of MIMO NFC from a DoF perspective. We began by
elucidating the spatial DoFs achievable in near-field SPD-MIMO and exploring how these increased DoFs can be exploited for enhancing the channel capacity. Next, we analyzed and compared three DoF-related performance metrics, namely ${\mathsf{EDoF}}_1$, ${\mathsf{EDoF}}_2$, and ${\mathsf{EDoF}}_3$, to their far-field counterparts for demonstrating the superiority of NFC in terms of spatial multiplexing and channel capacity. To further explore the potential of MIMO NFC, we extended these results to CAP-MIMO to determine the upper limit of performance. We have deepened the understanding of the augmented spatial DoFs offered by the near-field effect, with the hope of inspiring further innovations in this field. There are still numerous open research problems in this area, which are summarized from three aspects.
%However, it is important to acknowledge that there are still numerous open research problems in this area, which we have summarized into three main aspects.
\begin{itemize}
  \item DoF-Based Information-Theoretic Limits: The DoF is a significant information-theoretic measure directly related to channel capacity. Exploring the DoF to characterize the fundamental information-theoretic limits of NFC, including deriving the achievable DoF region, can provide essential insights for system design. Additionally, the pursuit of capacity-approaching transmission schemes for NFC from a DoF perspective represents a valuable endeavor.
\item DoF-Based Performance Analysis: Although our analysis has concentrated on point-to-point MIMO NFC, extending our investigations to multiuser scenarios holds the potential of offering valuable insights into the spatial DoFs in more complex communication setups, presenting a promising avenue for future research. Additionally, the heuristic nature of ${\mathsf{EDoF}}_2$ and the computational challenges in calculating ${\mathsf{EDoF}}_3$ for CAP-MIMO necessitate further research efforts to derive precise physical interpretations and practical implications for these metrics.
\item DoF-Inspired Beamforming Design: Effective beamforming designs are crucial for fully harnessing the increased DoFs offered by NFC. However, the computational and hardware complexities, particularly in the context of CAP-MIMO implementation, pose significant challenges. Therefore, there is a pressing need to explore scalable and computation-hardware efficient beamforming techniques that can exploit the benefits of augmented DoFs in practical NFC scenarios. %This research direction holds paramount importance for realizing the full potential of NFC technology in real-world applications.
\end{itemize}

\end{document}